\documentclass[11pt,a4paper]{article}
\usepackage{jheppub}
\usepackage{rotating}

\usepackage{amssymb}

\usepackage{graphics,bm}
\usepackage{epsfig}
\usepackage{graphicx}

\newcommand{\beq}{\begin{equation}}
\newcommand{\eeq}{\end{equation}}
\newcommand{\bqa}{\begin{eqnarray}}
\newcommand{\eqa}{\end{eqnarray}}

\def\sumint{\hbox{$\sum$}\!\!\!\!\!\!\int}
\def\square{\vcenter{\vbox{\hrule height.4pt
          \hbox{\vrule width.4pt height4pt
          \kern4pt\vrule width.3pt}\hrule height.4pt}}}

\title{Chiral perturbation theory in a magnetic background -
finite-temperature effects}

\author[a,b]{Jens O. Andersen}
\emailAdd{andersen@tf.phys.ntnu.no}
\date{Today}
\affiliation[a]{Department of Physics, 
Norwegian University of Science and Technology, 
H{\o}gskoleringen 5,
N-7491 Trondheim, Norway}
\affiliation[b]{
Niels Bohr International Academy, Niels Bohr Institute and Discovery Center,
Blegdamsvej 17, DK-2100 Copenhagen, Denmark}

\abstract{
We consider chiral perturbation theory for $SU(2)$
at finite temperature $T$ in a 
constant magnetic
background $B$. 
We compute the thermal mass of the pions 
and the pion decay constant to leading order in chiral perturbation
theory in 
the presence of the magnetic field.
The magnetic field gives rise to a splitting between $M_{\pi^0}$
and $M_{\pi^{\pm}}$ as well as between $F_{\pi^0}$ and $F_{\pi^{\pm}}$.
We also calculate the 
free energy and the quark condensate to next-to-leading order
in chiral perturbation theory.
Both the pion decay constants and the quark condensate are decreasing
slower as a function of temperature 
as compared to the case with vanishing magnetic field.
The latter result suggests
that the critical temperature $T_c$ for the chiral transition is larger
in the presence of a constant magnetic field.
The increase of $T_c$ as a function of $B$ is
in agreement with most
model calculations but in disagreement with recent lattice
calculations.}


\begin{document}
\maketitle
\section{Introduction}
While the QCD Lagrangian in the limit of zero quark masses has chiral
symmetry, the true ground state of QCD does not respect this symmetry.
Chiral symmetry is broken spontaneously in the vacuum by quantum effects.
Specifically, the $SU(N_f)_L\times SU(N_f)_R$ symmetry
of ${\cal L}_{\rm QCD}$
is broken down to $SU(N_f)_V$, and according to Goldstone's theorem
this gives rise to as many massless spin-zero particles as there are broken
generators. For $N_f$ fermions, this implies $N_f^2-1$ massless excitations
and in phenomenological, we have
applications $N_f=2$ or $N_f=3$. The symmetry breaking is
apparent in the low-energy spectrum of QCD 
from the existence of the three
very light pions. The fact that the pions are not strictly massless is
due to the fact that chiral symmetry of nature  is only 
approximate. The associated explicit symmetry breaking 
in QCD is built in by finite quark masses
in ${\cal L}_{\rm QCD}$. 

Lattice simulations are currently the only way to calculate the properties
of QCD from first principles. 
However, due to the sign problem, one is restricted to small values
of the baryon chemical potential.
Other approaches involve low-energy effective
theories or model calculations that share some of the properties of QCD
such as the same pattern of symmetry breaking. Examples are 
chiral perturbation theory (ChPT), the Nambu-Jona-Lasinio 
(NJL) model, linear sigma models, and quark-meson (QM) models.
At long wavelengths where the relevant degrees of freedom are the 
almost massless 
Goldstone bosons, one would naturally like to have an effective low-energy 
theory that contains only these degrees of freedom. Such an effective 
field theory may simplify calculations of the long-distance
properties of the system and one can obtain
model-independent predictions with a minimum of assumptions.

An example of a low-energy effective theory for QCD
is chiral perturbation theory~\cite{vinbjerg,gasser1, gasser12,bijn1}. 
The chiral Lagrangian that describes the (pseudo)Goldstone bosons
is uniquely determined by the global symmetries of QCD and the
assumption of chiral symmetry breaking. The Lagrangian ${\cal L}_{\rm eff}$
consists of a string of operators that involve an increasing number
of derivatives or quark mass factors, each multiplied by a low-energy
constant (LEC) $l_i$. 
However, QCD is a confining and strongly interacting
theory at low energies. Thus the coefficients $l_i$ of the chiral
Lagrangian cannot be calculated directly 
from QCD. Instead, the
coefficients of the effective Lagrangian are fixed by experiments.

Chiral perturbation theory provides a systematic framework at low energies.
It is not an expansion in powers of some small coupling constant, but it
is a a systematic expansion in powers of momenta $p$ where a derivative
counts as one power and the quark masses count as two powers~\cite{gasser1}.
Chiral perturbation theory is a nonrenormalizable quantum field theory in the 
old sense of the word. This means that a calculation at a given order $n$
in momentum $p$, requires that one adds higher-order operators in order
to cancel the divergences that arise in the 
calculations at order $n$. This implies
that one needs more and more couplings $l_i$ as one goes to higher 
loop orders, and therefore more experiments
to determine them. However, this poses no problem, as long as one is
content with finite precision. This is is the essence of effective
field theory~\cite{georgi,lepage,cliff}. 
The order $p^4$-divergences in ChPT were determined
in Ref.~\cite{gasser1}, while the order-$p^6$ divergences were
calculated in Ref.~\cite{bijn1}.

The thermodynamics of a pion gas using ChPT was studied in detail in 
a series of papers 25 years 
ago~\cite{gasser2,gasser3,gerber}. 
The temperature-dependence of the pion mass $M_{\pi}$ and the pion decay
constant $F_{\pi}$
were calculated to leading order (LO), while the
the pressure
and the temperature dependence of the quark condensate were calculated
to next-to-next-to-leading order (NNLO) in chiral
perturbation theory. 
In the chiral limit, there are two scales at finite temperature, namely
the pion decay constant and $T$, and chiral perturbation theory
is then an expansion in powers of $T^2/F_{\pi}^2$.
In the present paper, we give the calculational details of a number
of quantities in the presence of a constant magnetic field $B$.
These include $M_{\pi^0}$, $M_{\pi^{\pm}}$, $F_{\pi^0}$, $F_{\pi^{\pm}}$
at leading order and the free energy and the (normalized)
quark condensate at next-to-leading
order (NLO).


QCD in external magnetic fields has received a lot of attention in recent
years.
This is not a purely academic question since
the properties of QCD in strong magnetic fields $B$ is relevant in several 
situations. For example, 
large magnetic, $B\sim 10^{14}-10^{15}$ Gauss, exist inside
magnetars~\cite{duncan,neutron}. 
If the density of the core is sufficiently high, it contains 
quark matter.
In that case, the core
may be color superconducting and so it is important to study the effects
of external magnetic fields in this 
phase~\cite{mark,gorbar,qcdmag1,qcdmag2,fh,jorge,chinese,qcdmag3}.  
Similarly, it has been suggested that strong magnetic fields are created
in (noncentral)
heavy-ion collisions at the Relativistic Heavy-Ion Collider (RHIC) and 
the Large Hadron Collider (LHC) and that these play an important 
role~\cite{harmen1,rus1,rus2}.
In this case, the magnetic field  strength 
has been estimated to be up to $B\sim 10^{19}$ Gauss, which
corresponds to $|qB|\sim6m_{\pi}^2$, where $|q|$ is the charge of the pion.
Even larger fields could be reached due to the
effects of event-by-event fluctuations, see for example~\cite{china}.


This has spurred the interest in studying QCD in external fields.
At zero baryon chemical potential this can be done from first principles
using lattice simulations. Recently, lattice simulations in a constant
magnetic background $B$
have been carried out~\cite{sanf,negro, budaleik}.
At finite $\mu_B$ this is very difficult
due to the infamous sign problem. Therefore one often resorts to
effective theories that share some of the features of QCD, such as 
chiral symmetry breaking. 

A low-energy effective theory that provides a systematic
framework for systematic calculations is chiral perturbation theory.
Chiral perturbation theory has been used to study the
quark condensate in strong magnetic fields at zero 
temperature~\cite{smilga,shusp,cptB,werbos} and finite 
temperature~\cite{agacond}.
In Ref.~\cite{agam}, the thermal corrections to $M_{\pi^0}$ and
$F_{\pi^0}$ were computed, while the quark-hadron phase transition
was investigated in Ref.~\cite{chiralB}.
The effects of external magnetic fields 
on the chiral transition have been studied in detail
using the NJL model~\cite{klev,shovkovy+,gorbie,klim,hiller,boomsma2,chat,avan,frasca,rabbi},
the Polyakov-loop extended NJL model~\cite{pnjlgat,pnjlkas},
the QM model~\cite{frasca,rabbi,fraga1,rashid,anders}, 
the linear sigma model~\cite{duarte},
the (P)QM model~\cite{fragapol,skokov},
and the MIT bag model~\cite{fragamit}.


In the present paper, we study pions at finite 
temperature in a constant external magnetic field $B$ using 
chiral perturbation theory and it 
is organized as follows. In Sec.~II, we briefly discuss ChPT
in an external magnetic field. In Sec.~III, we calculate
the leading correction to the pion mass as well as the pion decay
constant. In Sec.~IV, we calculate the free energy and the
quark condensate to next-to-leading order in chiral perturbation theory.
In Sec.~V the numerical results are presented and we summarize in Sec.~VI.
The necessary sum-integrals are listed in Appendix A, while explicit 
calculations of those sum-integrals are presented in Appendix B.

\section{Chiral perturbation theory}
As explained in the introduction, chiral perturbation theory is
a low-energy effective field theory that can be used to
systematically calculate physical quantities as a power series in momentum
$p$.
The effective Lagrangian is given by an infinite string
of operators involving an increasing number of derivatives or quark masses.
Schematically, we can write
\bqa
{\cal L}_{\rm eff}&=&{\cal L}^{(2)}+{\cal L}^{(4)}
+{\cal L}^{(6)}
+...
\eqa
where the superscript indicates the powers of momentum. 
In Euclidean space, the leading term is given by
\bqa
{\cal L}^{(2)}&=&{1\over4}F^2{\rm Tr}
\left[(D_{\mu}U)^{\dagger}(D_{\mu}U)
-M^2{\rm Tr}(U+U^{\dagger})
\right]
\;.
\label{lo}
\eqa
The first term is the Lagrangian of the nonlinear sigma model and the
second term is the kinetic term for the gt
Here $U=\exp[i\tau^a\pi^a/F]$ is a unitary $SU(2)$ matrix, where
$\pi^a$ are the pion fields and $\tau^a$ are the Pauli spin matrices.
The low-energy constants $M$ and $F$ are the
tree-level values for the pion mass $M_{\pi}$ and the pion decay 
constant $F_{\pi}$, respectively.
Moreover $D_{\mu}$ is the covariant derivative which replaces 
the partial derivative:
\bqa
D_{\mu}&=&\partial_{\mu}-i[U,v_{\mu}]-i\left\{U,a_{\mu}\right\}
\;,
\eqa
where $v_{\mu}$ is an external vector field and $a_{\mu}$ is an 
external axial vector field.
The next-to-leading order terms in the low-energy expansion 
are~\cite{gasser1,gerber,werbos}
\footnote{The last two operators of Eq.~(\ref{p44}) are so-called contact terms.
The term $-h_1M^4$ contributes to the vacuum energy.
Loop corrections to the vacuum energy are divergent and 
renormalizing the couplings such as $h_1$ eliminates (some of)
the divergences and is needed to render the vacuum energy finite.
We will not need them in the following since we
are subtracting the divergences in the vacuum diagrams
that appear for $B=0$. The remaining divergences are then eliminated
by renormalizing the low-energy couplings $\bar{l}_i$.}
\bqa\nonumber
{\cal L}^{(4)}&=&-{1\over4}
l_1{\rm Tr}\left[(D_{\mu}U)^{\dagger}(D_{\mu}U)
\right]^2-
{1\over4}l_2{\rm Tr}
\left[(D_{\mu}U)^{\dagger}(D_{\nu}U)\right]
{\rm Tr}\left[(D_{\mu}U)^{\dagger}(D_{\nu}U)\right]
\\ && \nonumber
+{1\over8}l_4M^2
{\rm Tr}\left[(D_{\mu}U)^{\dagger}(D_{\mu}U)\right]
{\rm Tr}(U+U^{\dagger})
-{1\over16}(l_3+l_4)
M^4{\rm Tr}[(U+U^{\dagger})]^2
-l_5{\rm Tr}[U^{\dagger}F_{\mu\nu}^RUF_{\mu\nu}^L]
\\ &&
-{1\over2}il_6{\rm Tr}
[F_{\mu\nu}^R(D_{\mu}U)(D_{\nu}U^{\dagger})
+F_{\mu\nu}^L(D_{\mu}U^{\dagger})(D_{\nu}U)]
-h_1M^4
+h_2{\rm Tr}[
F_{\mu\nu}^lF_{\mu\nu}^l]
\label{p44}
\;.
\eqa
Here $F^l_{\mu\nu}=\partial_{\mu}F^l_{\nu}-\partial_{\nu}F^l_{\mu}
-i[F_{\mu}^l,F_{\nu}^l]$, where $l=L,R$, and 
$F_{\mu}^R=v_{\mu}+a_{\mu}$ and $F_{\mu}^L=v_{\mu}-a_{\mu}$.

The Lagrangian ${\cal L}^{(6)}$ is very complicated as it contains
more than 50
terms for $SU(2)$~\cite{bijn1}. However, only one term is relevant for the
calculations~\cite{shusp,cptB,werbos}, namely
\bqa
{\cal L}^{(6),\rm relevant}&=&-4c_{34}M^2(qF_{\mu\nu})^2\;.
\eqa
We consider in the following a constant external 
magnetic field so $v_{\mu}$ is an Abelian gauge field, 
$v_{\mu}={1\over2}qA_{\mu}\tau^3$, and $a_{\mu}=0$. 
Here $q$ is the electric charge of the 
pion and we choose the four-vector potential to be $A_{\mu}=B\delta_{\mu 2}x_1$.
By expanding the
Lagrangian Eq.~(\ref{lo})
to fourth order in the pion field $\pi$, we obtain
\bqa\nonumber
{\cal L}^{(2)}&=&-F^2M^2+
{1\over2}\left(\partial_{\mu}\pi^0\right)^2
+{1\over2}M^2\left(\pi^0\right)^2
+(\partial_{\mu}+iqA_{\mu})\pi^+(\partial_{\mu}-iqA_{\mu})\pi^-
+M^2\pi^+\pi^-
\\ &&\nonumber
-{M^2\over24F^2}\left[
(\pi^0)^2+2\pi^+\pi^-
\right]^2
+{1\over6F^2}\bigg\{
2\pi^0[\partial_{\mu}\pi^0][\partial_{\mu}(\pi^+\pi^-)]
-2\pi^+\pi^-(\partial_{\mu}\pi^0)^2
 \\&& 
-2\left[\left(\pi^0\right)^2+2\pi^+\pi^-\right]
(\partial_{\mu}\pi^+)(\partial_{\mu}\pi^-)
+[\partial_{\mu}(\pi^+\pi^-)]^2
\bigg\}
\;,
\label{trunc}
\eqa
where we have defined the complex pion fields
as $\pi_{\pm}={1\over\sqrt{2}}(\pi_1\pm i\pi_2)$
In the same manner, we can expand ${\cal L}^{(4)}$:
\bqa\nonumber
{\cal L}^{(4)}
&=&
{1\over4}F_{\mu\nu}^2+
{2l_5\over F^2}(qF_{\mu\nu})^2\pi^+\pi^-
+{2il_6\over F^2}qF_{\mu\nu}\left[
(\partial_{\mu}\pi^-)(\partial_{\nu}\pi^+)
+iqA_{\mu}\partial_{\nu}(\pi^+\pi^-)
\right]
+(l_3+l_4){M^4\over F^2}\left(\pi^0\right)^2
\\ && 
+2(l_3+l_4){M^4\over F^2}\pi^+\pi^-
+l_4{M^2\over F^2}(\partial_{\mu}\pi^0)^2+
2l_4{M^2\over F^2}
(\partial_{\mu}+iqA_{\mu})\pi^+
(\partial_{\mu}-iqA_{\mu})\pi^-\;.
\label{l44}
\eqa
We note in passing that our expression~(\ref{trunc}) for the
truncated Lagrangian ${\cal L}^{(2)}$ differs from the expression found
in Refs.~\cite{shusp,cptB} since they use a different parametrization
for the unitary matrix $U$, namely the Weinberg parametrization.
However, we obtain the same expressions for physical quantities
independent of parametrization~\cite{kapustelsen}. This is simply the statement
that physical quantities are independent of the coordinate system used.

At this point it is appropriate to briefly discuss the number
of Goldstone modes in the presence of an external electromagnetic field.
Due to the different electric charges of the $u$ and $d$ quarks,
flavor symmetry is broken in an external electromagnetic field.
In particular, the axial $SU(2)_A$-symmetry is broken down to a 
$U(1)_A^3$-symmetry, which corresponds to a rotation of the $u$
and $d$-quarks with opposite angles. The formation of a quark condensate
breaks this Abelian symmetry which gives rise to a Goldstone boson -
the neutral pion $\pi^0$~\cite{smilga}.
The charged pions are thus no longer Goldstone modes. In fact, the
presence of the external field allows for an effective mass term
even in the chiral limit when $M=0$, namely
the first two terms in Eq.~(\ref{l44}).

The chiral Lagrangian comes with a number of undetermined parameters or
low-energy constants (LECs). These parameters can be determined by 
experiments. However, loop corrections involve renormalization of
these parameters and the physical quantities are no longer
equal to the parameters of the chiral Lagrangian.
For example, $F$ and $M$ are no longer the 
measured pion decay constant and the physical pion mass, respectively.
The relation between the bare and renormalized parameters has been 
found in~\cite{gasser1} and can be expressed as~\footnote{In contrast to 
ref.~\cite{gasser1}, we have introduced the renormalization scale $\Lambda$
in the definition of the sum-integrals instead of as a part of the relation
between the bare and renormalized parameters. See Appendix A
for details.}
\bqa
l_i&=&-{\gamma_i\over2(4\pi)^2}\left[
{1\over\epsilon}+1-\bar{l}_i
\right]\;,
\label{normali}
\eqa
where $\gamma_i$ are coefficients that are tabulated in~\cite{gasser1} 
and $\bar{l}_i$ are scale-independent parameters, i. e. they are the
renormalized running couplings evaluated
at the renormalization scale $\Lambda=M$. 

In the present calculations, we
need $\gamma_3=-{1\over2}$, $\gamma_4=2$, $\gamma_5=-{1\over6}$, and 
$\gamma_6=-{1\over3}$~\cite{gasser1,gasser12}. 
The coupling $c_{34}$ in ${\cal L}^{(6)}$
can be renormalized as in~\cite{bijn1} by using
\bqa
c_{34}
&=&{1\over F^2}
\left[
\bar{c}_{34}-{1\over2(4\pi)^2\epsilon}\left(
\bar{l}_5-{1\over2}\bar{l}_6
\right)
\right]\;.
\eqa
We close this section by briefly discussion the bare propagators
of the pions in a magnetic backround.
The classical solutions to the Klein-Gordon equation in a constant magnetic
field $B$ are well known and the dispersion relation is given by
\bqa
\left(E_{m,p_z}^{\pm}\right)^2&=&p_z^2+M^2+(2m+1)|qB|\;,
\eqa
where 
$m=0,1,2,...$ denotes the $m$th Landau level, 
$q$ is the electric charge of the pion,
and $p_z$ is the spatial momentum
in the $z$-direction. The subscript $\pm$ denotes $\pi^{\pm}$ 
and we note that the dispersion relations for the two charged
pions are identical.
The imaginary-time propagator for a charged pions is then
$\Delta_{\pi^{\pm}}(P_0,{\bf p})=1/(P_0^2+p_z^2+M^2+(2m+1)|qB|)$, while 
for neutral pions it
is given by the usual 
$\Delta_{\pi^0}(P_0,{\bf p})=1/(P_0^2+p^2+M^2)$, where $P_0=2n\pi T$
is the $n$th Matsubara frequency.

\bibliographystyle{elsarticle-num}
\bibliography{<your-bib-database>}







\section{Pion masses and pion decay constants}
In this section, we discuss the thermal pion masses and the pion decay
constants. In order to calculate the thermal pion masses, we consider
the inverse propagators at one-loop. 
For example, the inverse propagator for the neutral pion can be written as
\bqa
\Delta^{-1}_{\pi_0}(P_0,{\bf p})&=&
P^2+M^2+\Sigma_{\pi_0}(P_0,{\bf p})\;, 
\eqa
where the self-energy is given by
\bqa
\Sigma_{\pi_0}(P_0,{\bf p})&=&
[A_{0}-\Delta Z_{0}]P^2+[B_{0}-\Delta Z_{0}]M^2
\;, 
\eqa
where $\Delta Z_0$ 
is
the wavefunction renormalization counterterm
for $\pi^0$.
 and $\pi^{\pm}$, respectively.
The coefficients $A_0$ and $B_0$ 
are given by 
\bqa
A_0&=&
-{2\over3F^2}\sumint_P^B{1\over P_0^2+p_z^2+M^2_B}
+2{M^2\over F^2}l_4\;,
\label{w1}
\\
B_0&=&
{1\over6F^2}\left[2\sumint_P^B{1\over P_0^2+p_z^2+M^2_B}
-3\sumint_P{1\over P_0^2+p^2+M^2}\right]
+2{M^2\over F^2}(l_3+l_4)
\label{cont1}
\;.
\eqa
The sum-integrals $\sumint_P$ and $\sumint_P^B$
are defined in Appendix A.
The counterterms $\Delta Z_0$ 
is 
chosen such that the self-energy
$\Sigma_{\pi_0}$ 
is independent of the momentum of the pions~\cite{loewe}. 
This implies that $\Delta Z_0=A_0$.
The propagator then reduces to
\bqa
\Delta^{-1}_{\pi_0}(P_0,{\bf p})&=&
P^2+M^2\left[1-A_0+B_0\right]\;.
\label{proppie}
\eqa
Inserting Eqs.~(\ref{w1}) and~(\ref{cont1}) into
Eq.~(\ref{proppie}),
we can then read off the mass $m_{\pi^0}$ from the propgator and we find
\bqa
M_{\pi_0}^2&=&M^2
+{M^2\over F^2}\sumint_P^B{1\over P_0^2+p_z^2+M_B^2}
-{1\over2}{M^2\over F^2}
\sumint_P{1\over P^2+M^2}
+2l_3{M^4\over F^2}
\;.
\eqa
The mass $M_{\pi^{\pm}}$ of the charged pion can
be found in a similar way
\bqa
M_{\pi^{\pm}}^2&=&M^2
+{1\over2}{M^2\over F^2}\sumint_P{1\over P^2+M^2}
+2l_3{M^4\over F^2}+{2\over F^2}(2l_5-l_6)(qB)^2
\label{mpippp}
\;.
\eqa
Renormalization of $l_5$ and $l_6$ are carried out
according to Eq.~(\ref{normali}), which shows
that we can replace
$2l_5-l_6$ by $(\bar{l}_6-\bar{l}_5)/6(4\pi)^2$~\cite{shusp,cptB}.
After renormalization, one obtains
\bqa
M_{\pi^0}^2&=&M^2_{\pi}\bigg[1-
{1\over(4\pi)^2F^2}\bigg(
I_B(M)
+{1\over2}J_1(\beta M)T^2-J_1^B(\beta M)|qB|\bigg)
\bigg]\;,
\label{mpi0}
\\
M^2_{\pi^{\pm}}&=&M^2_{\pi}\left[1+{1\over2(4\pi)^2F^2}J_1(\beta M)T^2\right]
+{(qB)^2\over3(4\pi)^2F^2}(\bar{l}_6-\bar{l}_5)
\;,
\label{mpip}
\eqa
where the function $I_B(M)$ is defined by
\bqa\nonumber
I_B(M)&=&
M^2\log{M^2\over2|qB|}-M^2-2\zeta^{(1,0)}(0,\mbox{$1\over2$}+x)|qB|
\\ &=&
M^2\log{M^2\over2|qB|}-M^2-2\log[\Gamma(0,\mbox{$1\over2$}+x)]|qB|
+\log(2\pi)|qB|
\label{ibdef}
\;,
\eqa 
and $x={M^2\over2|qB|}$. In particular, note that $I_B(0)=|qB|\log2$.
The thermal integrals 
$J_n(\beta M)$ and $J_n^B(\beta M)$ are defined in Appendix A, 
and evaluated at $\epsilon=0$.
Moreover, 
$\zeta^{(1,0)}(-b,y)={d\over dx}\zeta(x-b,y)|_{x=0}$, where $\zeta(s,x)$
is the Hurwitz zeta-function and $b$ is a real number.
The physical pion mass $M_{\pi}^2$ in the vacuum is given by 
\bqa
M_{\pi}^2&=&M^2\left[
1-
{M^2\over2(4\pi)^2F^2}\bar{l}_3
\right]\;.
\label{pit0}
\eqa
The result~(\ref{mpi0}) is in agreement with
the calculations of Ref.~\cite{agam}.
In the limit $B\rightarrow0$, Eqs.~(\ref{mpi0})--(\ref{mpip})
reduce to the result of Ref.~\cite{gasser2}.
We notice that temperature dependence of the charged pion mass $M_{\pi^{\pm}}$
is the same as 
for vanishing magnetic field since the only loop correction to the
mass involves a neutral pion, cf. Eq.~(\ref{mpippp}).
The only difference is a temperature-independent
constant proportional to $(qB)^2$. This constant survives the 
the chiral limit, 
$M\rightarrow0$, which shows that the charged pion is no longer a Goldstone
mode in the presence of an electromagnetic field.

We next discuss the pion decay constant $F_{\pi^0}$ for the neutral 
pion. 
The components of the 
axial current
${\cal A}_{\mu}^0$ are given by~\cite{kapustelsen,loewe}
\bqa
{\cal A}^0_{\mu}&=&-F\partial_{\mu}\pi^0
+{2\over3F}\left[
2\pi^+\pi^-\partial_{\mu}\pi^0-\pi^0\partial_{\mu}(\pi^+\pi^-)
\right]
-2{M^2\over F}l_4\partial_{\mu}\pi^{0}\;,
\label{a0}
\eqa
In order to calculate the pion decay constant, we need to evaluate the
matrix element
$F_{\pi^0}=\langle0|{\cal A}_{\mu}^0|\pi^{0}\rangle$.
The matrix elements are proportional to $ip_{\mu}$.
Moreover, in the leading term $-F\partial_{\mu}\pi^0$
of Eq.~(\ref{a0}), 
we must take into account wavefunction
renormalization of the pion field
given by Eqs.~(\ref{w1}). 
A straightforward
calculation of the matrix element then gives
\bqa
F_{\pi^0}&=&
F\left[1-{1\over F^2}\sumint_P^B{1\over P_0^2+p_z^2+M_B^2}
+{M^2\over F^2}l_4\right]\;
\eqa
In a similar way we can calculate the pion decay constant for the 
charged pion, $F_{\pi^{\pm}}$, and one finds
\bqa
F_{\pi^{\pm}}
&=&
F\left[1-{1\over2F^2}\sumint_P{1\over P^2+M^2}
-{1\over2F^2}\sumint_P^B{1\over P_0^2+p_z^2+M_B^2}
+{M^2\over F^2}l_4\right]\;.
\eqa
After renormalization, we find
\bqa
\label{fpi00}
F_{\pi^0}&=&F_{\pi}\left[1
+{1\over(4\pi)^2F^2}
\left(
I_B(M)-
J_1^B(\beta M)|qB|\right)\right]\;, \\ 
F_{\pi^{\pm}}&=&F_{\pi}\bigg[1+{1\over2(4\pi)^2F^2}
\bigg(
I_B(M)
-J_1(\beta M)T^2-J_1^B(\beta M)|qB|\bigg)\bigg]\;,
\label{fpipm0}
\eqa
where the thermal integrals $J_1(\beta M)$ and $J_1^B(\beta M)$ 
are evaluated at $\epsilon=0$,and
where
the pion decay constant $F_{\pi}$ in the vacuum is 
\bqa
F_{\pi}&=&\left[
1+{M^2\over(4\pi)^2F^2}\bar{l}_4
\right]\;.
\eqa
The result~(\ref{fpi00}) agrees with the calculations of
Agasian and Shushpanov~\cite{agam}.
In the limit $B\rightarrow0$, $F_{\pi^0}=F_{\pi^{\pm}}$ and they reduce to
the result of~\cite{gasser3} as they should.
At zero magnetic field, the decay constants are identical 
$F_{\pi^o}=F_{\pi^{\pm}}$ but there are 
two pion decay constants at finite temperature;
one for the
time component and one for the spatial component of ${\cal A}_{\mu}$.
The difference between them is an order-$p^4$ effect~\cite{pisarskipions}
and this the calculation of the diffference is beyond the scope of this paper.

\section{Free energy  and quark condensate}
In this section, we compute the free energy and the quark condensate
to NLO in chiral perturbation theory. 
We are interested in the contributions to the free energy
that are due to a nonzero magnetic field and finite temperature.
We therefore write the contribution to the free energy 
at the $n$th loop order, ${\cal F}_n$,  as
a sum of three different terms 
\bqa
\label{totalf}
{\cal F}_n&=&{\cal F}^{\rm vac}_n+{\cal F}^{B}_n+{\cal F}^{T}_n\;,
\eqa
where ${\cal F}^{\rm vac}_n$ is the contribution in the vacuum, i. e. 
for $B=T=0$, ${\cal F}^{B}_n$ is the zero-temperature contribution due
to a finite magnetic field, and ${\cal F}^{T}_n$ is the finite-temperature
contribution. The last term also depends on $B$.
The strategy is to isolate the term
${\cal F}^{\rm vac}_n$ and subtract it from Eq.~(\ref{totalf}).
This term contains ultraviolet divergences which are removed by 
renormalization of the low-energy constants of the chiral Lagrangian
in the usual way and the renormalized expression for 
${\cal F}_n^{\rm vac}$ represents the contribution to the 
vacuum energy of the theory from the $n$th loop order.
The term ${\cal F}^{B}_n$ generally contains additional
ultraviolet divergences as well and it is rendered finite
by the same renormalization procedure.
The term ${\cal F}^{T}_n$ is finite.

The free energy at tree level is given by
the first term in Eqs.~(\ref{trunc}) and~(\ref{l44}):
\bqa
{\cal F}_0&=&{1\over2}B^2-F^2M^2\;.
\label{treef}
\eqa
The first term in Eq.~(\ref{treef}) is needed 
when we renormalize the one-loop contribution to 
the free energy in the next subsection.
The second term is independent
of $B$, so we neglect it in the following.

\subsection{One-loop free energy}
The one-loop contribution ${\cal F}_1$ to the free energy 
can be written as
the sum of the contributions from the charged and neutral pions, and
tree-level terms from ${\cal L}^{(4)}$
\bqa
{\cal F}_1&=&
{\cal F}_{\pi^0}
+{\cal F}_{\pi^{+}}+{\cal F}_{\pi^{-}}
+{\rm tree\,\,level\,\,terms\,\,from}\,\,{\cal L}^{(4)}
\;,
\eqa
where
\bqa
{\cal F}_{\pi^0}&=&{1\over2}\sumint_P\log\left[{P^2+M_{\pi}^2}\right]\;, \\
{\cal F}_{\pi^{\pm}}&=& 
{1\over2}
\sumint_P^{B}
\log\left[P_0^2+p_z^2+M^2_B\right]
\;,
\label{fpisum}
\eqa
The
tree-level terms are necessary in the renormalization of the vacuum
energy, but according to our strategy we do not need to consider these
since they are independent of the magnetic field.
Using the expressions for the various sum-integrals in Appendix A, we
obtain
\bqa\nonumber
{\cal F}_1&=&
{1\over2(4\pi)^2}
\left({\Lambda^2\over2|qB|}\right)^{\epsilon}
\left[
\left({(qB)^2\over3}-
M^4\right)\left({1\over\epsilon}+1\right)
+8
\zeta^{(1,0)}(-1,\mbox{$1\over2$}+x)(qB)^2
-2J_0^B(\beta M)|qB|T^2
\right]
\\ &&
-{1\over4(4\pi)^2}\left({\Lambda^2\over M^2}\right)^{\epsilon}
\left[
\left({1\over\epsilon}+{3\over2}\right)M^4
+2J_0(\beta M)T^4
\right]
\;.
\eqa
The divergence that depends on the external magnetic field $B$ is removed
by wavefunction renormalization, i. e. by the replacement
\bqa
B^2&\rightarrow&B^2\bigg[1-{q^2\over6(4\pi)^2\epsilon}
\bigg]\;,
\eqa
in the tree-level term ${1\over2}B^2$ in Eq.~(\ref{treef}). 
Subtracting the expressions for the diagrams for $T=B=0$, we remove
the remaining divergences and obtain a finite result.
One finds
\bqa
{\cal F}_1^{B}&=&
{M^4\over4(4\pi)^2}\left[1-2\log{M^2\over2|qB|}\right]
+{4(qB)^2\over(4\pi)^2}\zeta^{(1,0)}(-1,\mbox{$1\over2$}+x)
+{(qB)^2\over6(4\pi)^2}\log{\Lambda^2\over2|qB|}\;,
\label{f1bb}
\\ 
{\cal F}_1^{T}&=&
-{1\over2(4\pi)^2}\left[
J_0(\beta M)T^4+2J_0^B(\beta M)|qB|T^2
\right]
\label{1lv}\;.
\eqa

\subsection{Two-loop free energy }
\begin{figure}[htb]
\begin{center}
\includegraphics[width=14cm]{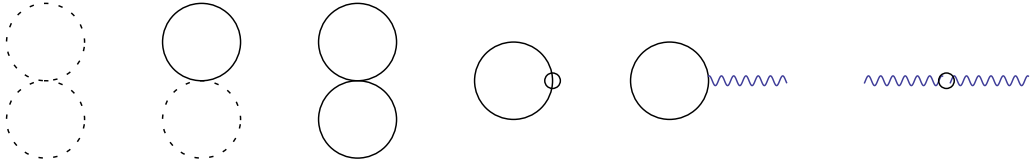}
\caption{Feynman diagrams contributing to the free energy at order $p^4$.}
\label{nlo}
\end{center}
\end{figure}

The order-$p^4$ diagrams that contribute to the free energy are shown in 
Fig.~\ref{nlo}. Dotted lines denote a neutral pion and solid lines
denote a charged pion. Wiggly line denote the background field $|qB|$.
The two-loop contribution ${\cal F}_2$
to the free energy can be written as
\bqa
{\cal F}_2&=&
{\cal F}_{2a}
+{\cal F}_{2b}+{\cal F}_{2c}+{\cal F}_{2d}+{\cal F}_{2e}
\;,
\eqa
where the expressions for the diagrams are~\footnote{It was shown
in Refs.~\cite{shusp,werbos} 
that the diagram ${\cal F}_{2c}$ vanishes
at $T=0$. It also vanishes at finite temperature.}
\bqa
{\cal F}_{2a}&=&
-{1\over8}{M^2\over F^2}\left(\sumint_P{1\over P^2
+M^2}\right)^2
\\
{\cal F}_{2b}&=&
{1\over2}{M^2\over F^2}
\sumint_P^B{1\over P^2+M_B^2}
\sumint_Q{1\over Q^2+M^2}
\\
{\cal F}_{2c}&=&
0 
\;,\\
{\cal F}_{2d}&=&
l_3{M^4\over F^2}\sumint_P{1\over P^2+M^2}
+2l_3{M^4\over F^2}\sumint_{P}^B{1\over P^2+M_B^2}\;, \\
{\cal F}_{2e}&=&
{2(qB)^2\over F^2}(2l_5-l_6)\sumint_P^B{1\over P^2+M_B^2}
\;,\\
{\cal F}_{2f}&=&
-8c_{34}(qB)^2M^2
\;.
\eqa
Using the expressions for the sum-integrals in Appendix A, we
obtain
\bqa\nonumber
{\cal F}_{2a}&=&
-{1\over8(4\pi)^4}{M^2\over F^2}\left({\Lambda^2\over M^2}\right)^{2\epsilon}
\bigg[
\left({1\over\epsilon^2}+{2\over\epsilon}+{\pi^2\over6}+3\right)M^4
-2\left({1\over\epsilon}+1\right)J_1(\beta M)M^2T^2
\\ &&
+J_1^2(\beta M)T^4
+{\cal O}(\epsilon)
\bigg]\;,
\\ \nonumber
{\cal F}_{2b}
&=& 
{1\over2(4\pi)^4}{M^2\over F^2}\left({\Lambda^2\over M^2}\right)^{\epsilon}
\left({\Lambda^2\over2qB}\right)^{\epsilon}
\left[
\left({1\over\epsilon^2}+{1\over\epsilon}
+{\pi^2\over6}+1\right)M^4
-2\left({1\over\epsilon}+1\right)\zeta^{(1,0)}(0,\mbox{$1\over2$}+x)|qB|
\right.\\ && \nonumber\left.
-\zeta^{(2,0)}(0,\mbox{$1\over2$}+x)|qB|M^2
-{1\over\epsilon}\left(J_1(\beta M)T^2+J_1^B(\beta M)|qB|\right)
M^2-J_1^B(\beta M)|qB|M^2
\right.\\ &&\left.
+J_1(\beta M)J_1^B(\beta M)T^2|qB|
+2\zeta^{(1,0)}(0,\mbox{$1\over2$}+x)J_1(\beta M)|qB|T^2
+{\cal O}(\epsilon)\right]\;,
\\
{\cal F}_{2c}&=&0\;,
\\ {\cal F}_{2d}
&=& 
-{l_3\over(4\pi)^2}{M^4\over F^2}\left({\Lambda^2\over M^2}\right)^{\epsilon}
\left\{
\left[
{1\over\epsilon}+1
+\left({\pi^2\over12}+1\right)\epsilon\right]M^2
-J_1(\beta M)T^2+{\cal O}(\epsilon^2)
\right\}\;,
\\ && \nonumber
-{2l_3\over(4\pi)^2}{M^4\over F^2}\left({\Lambda^2\over2|qB|}\right)^{\epsilon}
\left[
\left({1\over\epsilon}+{\pi^2\over12}\epsilon\right)M^2
-2\zeta^{(1,0)}(0,\mbox{$1\over2$}+x)|qB|
\right.\\&&\left.
-\zeta^{(2,0)}(0,\mbox{$1\over2$}+x)|qB|\,\epsilon
-J_1^B(\beta M)|qB|
+{\cal O}(\epsilon^2)
\right]\;,
\\ \nonumber
{\cal F}_{2e}
&=&
{2(qB)^2\over F^2}(2l_5-l_6)
{1\over(4\pi)^2}\left({\Lambda^2\over2qB}\right)^{\epsilon}
\bigg[
{1\over\epsilon}M^2
-2\zeta^{(1,0)}(0,\mbox{$1\over2$}+x)|qB|
-J_1^B(\beta M)|qB|
\\ &&
+{\cal O}(\epsilon)
\bigg]\;.
\eqa
After having subtracted the expressions for the diagrams for $T=B=0$, 
we are still left with some simple poles in $\epsilon$.
These remaining divergences are now elimated by renormalizing 
the couplings $l_3$, $l_5$, and $l_6$ according to the
prescription~(\ref{normali}). Since the renormalized couplings 
$l_i^r$
are evaluated
at the scale $\Lambda^2=M^2$, the final result simplies significantly.
After a lengthy calculations, we obtain
\bqa
{\cal F}^{B}_2&=&
{(qB)^2\over2(4\pi)^4F^2}\Bigg\{
I_B\left[
{M^4\over(qB)^2}
\bar{l}_3
-{2\over3}(\bar{l}_6-\bar{l}_5)\right]
-2\bar{d}M^2\Bigg\}\;,
\\ \nonumber
{\cal F}^T_2&=&
{M^2\over8(4\pi)^4F^2}\bigg[
-J_1^2(\beta M)T^4+4J_1(\beta M)J_1^B(\beta M)T^2|qB|
-2\bar{l}_3(J_1(\beta M)T^2+2J_1^B(\beta M)|qB|)M^2
\\ && 
-4I_BJ_1(\beta B)T^2
\bigg]
+{(qB)^2\over3(4\pi)^4F^2}\left(\bar{l}_6-\bar{l}_5\right)J_1^B(\beta M)|qB|
\;,
\eqa
where
\bqa
\bar{d}(M^2)&=&8(4\pi)^4c_{34}^r
+{1\over3}(\bar{l}_6-\bar{l}_5)\log{\Lambda^2\over M^2}
\;.
\eqa
Note in particular that all the divergent terms involving the
thermal integrals $J_1$ and $J_1^B$ cancel amongst themselves and so
we can evaluate the remaining ones at $\epsilon=0$.

If we express the contributions ${\cal F}^B_{1}$ 
and ${\cal F}^T_{1}$ in terms of
the physical pion masses $M_{\pi^0}(0)$ 
and $M_{\pi^{\pm}}(0)$ 
using Eqs.~(\ref{mpi0}) and~(\ref{mpip}) 
instead of the parameter$M$, the dependence on $\bar{l}_3$, $l_5$,
and $l_6$
cancels in the sums ${\cal F}_{1+2}^B$ and ${\cal F}_{1+2}^T$.
This yields
\bqa\nonumber
{\cal F}^{B}_{1+2}&=&
{M_{\pi^{\pm}}^4(0)\over4(4\pi)^2}
\left[1-2\log{M_{\pi^{\pm}}^2(0)\over2|qB|}\right]
+{4(qB)^2\over(4\pi)^2}\zeta^{(1,0)}(-1,\mbox{$1\over2$}+x_{\pm})
+{(qB)^2\over6(4\pi)^2}\log{\Lambda^2\over2|qB|}
\\ &&
-{(qB)^2\over(4\pi)^4F^2}
\bar{d}(|qB|)M
\Bigg]
\label{f2bb}
\;,
\\ \nonumber
{\cal F}^T_{1+2}&=&
-{1\over2(4\pi)^2}\bigg[J_0(\beta M_{\pi^0}(0))T^4
+2J_0^B(\beta M_{\pi^{\pm}}(0))|qB|T^2\bigg]
+{M^2\over8(4\pi)^4F^2}\bigg[
-J_1^2(\beta M)T^4
\\&&
+4J_1(\beta M)J_1^B(\beta M)T^2|qB|
-4I_BJ_1(\beta M)T^2
\bigg]
\label{f222} 
\;,
\eqa
where $x_{\pm}=M_{\pi^{\pm}}^2(0)/2|qB|$.
The result for the vacuum energy ${\cal F}_{1+2}^B$
is in agreement with the calculation
of Ref.~\cite{werbos}
~\footnote{Note that the final result of 
Ref.~\cite{werbos} is expressed in terms of the pion mass $M_{\pi}$
in the vacuum and so the terms involving $\bar{l_5}$ and $\bar{l}_6$
are not absorbed in the order $p^4$ term.}. 
Since $M_{\pi^{\pm}}$ is nonzero in the chiral limit ($M=0$)
and it appears in the thermal integrals $J_0^B$
in Eq.~(\ref{f222}),
the NLO correction to the
free energy does not vanish in contrast to 
the case of zero magnetic field.

We next discuss the quark condensate in the presence of the
the magnetic field $B$.
At finite temperature, the quark condensate is~\cite{gerber}
\bqa
\langle\bar{q}q\rangle
&=&
\langle0|\bar{q}q|0\rangle
\left[
1-{c\over F^2}{\partial({\cal F}^B+{\cal F}^{T})\over\partial M_{\pi}^2}
\right]\;,
\eqa
where $c$ is a constant defined by
\bqa
c&=&-F^2{\partial M_{\pi}^2\over\partial m}\langle0|\bar{q}q|0\rangle^{-1}
\;,
\eqa
where $m$ is the quark mass.
In the chiral limit, we have $c=1$~\cite{gerber}. 
In this case, the
quark condensate reduces to
\bqa\nonumber
\langle\bar{q}q\rangle
&=&
\langle0|\bar{q}q|0\rangle
\left\{
1+{|qB|\over(4\pi)^2F^2}
I_B(M_{\pi^{\pm}}(0))
+{(qB)^2\over(4\pi)^4F^4}
\bar{d}(qB)
-{1\over2(4\pi)^2F^2}\Big(J_1(0)T^2
\right. \\ &&\left. \nonumber
+2J_1^B(\beta M_{\pi^{\pm}}(0))|qB|\Big)
-{T^2\over8(4\pi)^4F^4}\left(
-J_1^2(0)T^2+4J_1(0)J_1^B(0)|qB|
-4\log2\,J_1(0)|qB|\right)
\right\}\;,
\\ &&
\label{quarkie}
\eqa
where we have used the recursion relations for $J_n$
and $J_n^B$.
We notice that we in the limit $B\rightarrow0$ recover the result
by Gerber and Leutwyler~\cite{gerber}
\bqa
\langle\bar{q}q\rangle
&=&
\langle0|\bar{q}q|0\rangle
\left[
1-{T^2\over8F^2}
-{T^4\over384F^4}
\right]\;,
\label{condb0}
\eqa
where we have used $J_1(0)=4\pi^2/3$.

\section{High-temperature expansion}
\label{numres}
In this section, we discuss the expansion of our results in the limit
of weak fields, i.e. for $|qB|/T^2\ll1$.
For simplicity, we restrict our analysis to the chiral limit, 
$M=0$.

For values $|qB|/T^2\leq1$, the 
sum over Landau levels in $J_n^B(\beta M)$
converges very slowly. It is advantageous to use the 
Euler-McLaurin summation formula~\cite{math} 
to evaluate $J_n^B$
in this regime~\cite{duarte}.
The sum over integers $m$ can be written as
\bqa\nonumber
\sum_{m=k}^{l}
f(m)
&=&
\int_k^lf(x)\,dx
+{1\over2}\left[
f(l)+f(k)
\right]
+\sum_{i=1}^{n}{b_{2i}\over(2i)!}\left[
f^{(2i-1)}(l)-f^{(2i-1)}(k)
\right]
\\ &&
+\int_k^l{B_{2n+1}(\{x\})\over(2n+1)!}f^{(2n+1)}(x)\,dx\;,
\label{euler}
\eqa
where $B_n$ are the Bernoulli numbers and $\{x\}$
means the fractional part of $x$. The last term in Eq.~(\ref{euler}) is the 
remainder. 
We will illustrate the use of Eq.~(\ref{euler}) in the case
of $J_1^B$.
Setting $n=2$, $k=0$, and $l=\infty$, we obtain
\bqa\nonumber
J_1^B
&=&8
\int_0^{\infty}dz
\left\{
\int_0^{\infty}
{dk\over E(z,M,T,B,k)}
{1\over e^{E(z,M,T,B,k)}-1}
+{1\over2}
{1\over E(z,M,T,B,0)(e^{E(z,M,T,B,0)}-1)}
\right.\\ && \left.
+{|qB|\over12T^2}
{1\over E^2(z,M,T,B,0)(e^{E(z,M,B,T,0)}-1)}
\left[
{1\over E(z,M,T,B,0)}+
{e^{E(z,M,T,B,0)}\over e^{E(z,M,T,B,0)}-1}
\right]
\right\}+...\;,
\label{j1b0}
\eqa
where we have defined 
$E(z,M,k,T)=\sqrt{z^2+M^2/T^2+(2k+1)|qB|/T^2}$. 
Integrating over $k$ in the first term of Eq.~(\ref{j1b0}), we obtain
\bqa\nonumber
J_1^B&=&
-{8\over a^2}
\int_0^{\infty}dz\,\log\left[
1-e^{-\sqrt{z^2+a^2}}
\right]
+4
\int_0^{\infty}
{dz\over\sqrt{z^2+a^2}}{1\over e^{\sqrt{z^2+a^2}}-1}
\\ &&
+{2a^2\over3}
\int_0^{\infty}
{dz\over z^2+a^2}{1\over e^{\sqrt{z^2+a^2}}-1}
\left[
{1\over\sqrt{z^2+a^2}}
+{e^{\sqrt{z^2+a^2}}\over e^{\sqrt{z^2+a^2}}-1}
\right]+...\;,
\label{qont}
\eqa
where $a^2=|qB|/T^2$. 
The Euler-McLaurin formula is next used to expand the integrals $J_n^B$
in a Taylor series
around $B=0$. 
This is particularly simple in the chiral limit where
$J_n^B$ only depends on the ratio $|qB|/T^2$, and 
the series is in powers of $\sqrt{qB}/T$. 
Integrating the first and the last term in Eq.~(\ref{qont}) by parts, 
we can write
\bqa
J_1^B&=&{1\over a^2}J_1
+J_2
+{a^2\over3}J_3
+...\;,
\eqa
where the argument of $J_n$ is $\beta a$.
The expansion of the integrals $J_n(x)$ can be found in e.g.~\cite{spt3}
and the leading terms are $J_1=4\pi^2/3-4\pi a$,
$J_2=2\pi/a$, and $J_3=\pi/a^3$.
This yields
\bqa
J_1^B|qB|&=&{4\pi^2T^2\over3}
\left[
1-{5\over4\pi}{\sqrt{qB}\over T}
+{\cal O}\left({qB\over T^2}\right)\right]\;.
\eqa
Inserting this expression into Eqs.
~(\ref{fpi00}), and~(\ref{fpipm0}), we obtain
in the chiral limit
\bqa
F_{\pi^0}&=&F_{\pi}\left[1+{|qB|\over(4\pi)^2F^2}\log2
-{T^2\over12F^2}+{5\sqrt{qB}T\over48\pi F^2}+...
\right]\;, \\ 
F_{\pi^{\pm}}&=&F_{\pi}\bigg[1
+{|qB|\over2(4\pi)^2F^2}\log2
-{T^2\over12F^2}+{5\sqrt{qB}T\over96\pi F^2}+...
\bigg]\;.
\eqa

In the same manner, we can expand the temperature-dependent part
of the quark condensate around $qB=0$. A straightforward calculation yields
\bqa
\langle\bar{q}q\rangle
&=&
\langle0|\bar{q}q|0\rangle
\left[
1-{T^2\over8F^2}+{5\sqrt{qB}T\over48\pi F^2}
-{T^4\over384F^4}
+{5\sqrt{qB}T^3\over1536\pi F^4}
+{T^2|qB|\over384F^4}\log2
+...\right]\;.
\eqa

\section{Numerical results and discussion}

We need to evaluate the integrals $J_0^B$ and $J_1^B$.
By expanding the Bose-Einstein distribution function in $J_0^B$
and $J_1^B$,
and integrating over three-momenta, we write it as a double sum 
involving a modified Bessel function $K_l(x)$ of order $l=0$: 
\bqa
\label{jb0k}
J_0^B(\beta M)&=&-8\sum_{n=1}^{\infty}\sum_{m=0}^{\infty}
\sqrt{M_{\pi}^2+(2m+1)|qB|\over T}
K_1\left(\mbox{$n\sqrt{M_{\pi}^2+(2m+1)|qB|}\over T$}\right)\;,\\
J_1^B(\beta M)&=&8\sum_{n=1}^{\infty}\sum_{m=0}^{\infty}
K_0\left(\mbox{$n\sqrt{M_{\pi}^2+(2m+1)|qB|}\over T$}\right)\;.
\label{jb1k}
\eqa
For values $|qB|/T^2\geq1$, the sums converge fast.
In the remainder of the paper, we use $|qB|=5(140 {\rm MeV})^2$
and so we use the expressions~(\ref{jb0k})--(\ref{jb1k})
in the numerical work.

\begin{figure}[htb]
\begin{center}
\setlength{\unitlength}{1mm}
{\includegraphics[width=8cm]{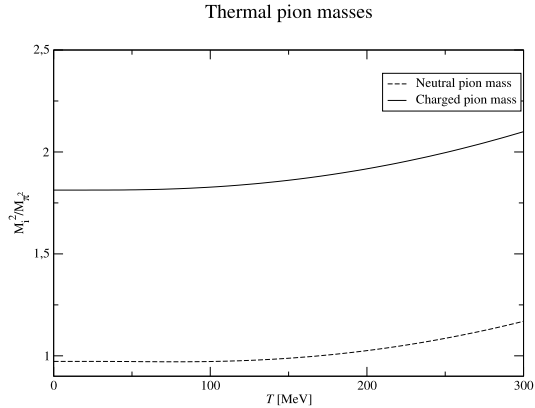}}
\caption{Pion masses $M_{\pi^0}^2$ and $M_{\pi^{\pm}}^2$
normalized to their vacuum
valued $M_{\pi}^2$ as a function of $T$.}
\label{thmass}
\end{center}
\end{figure}

In Fig.~\ref{thmass}, we show the
pion masses 
$M_{\pi^0}^2$ and $M_{\pi^{\pm}}^2$ given by Eqs.~(\ref{mpi0}) and~(\ref{mpip})
normalized to their vacuum
value $M_{\pi}^2$ as a function of $T$.
We use the experimental value
$F_{\pi}=93$ MeV. Moreover, the scale-independent difference 
of $\bar{l}_{6}$ and $\bar{l}_{5}$ is
$\bar{l}_6-\bar{l}_5=3\pm0.3$~\cite{belluci,bijn2}.
In the remainder we use central value of this difference, i e. 
$\bar{l}_6-\bar{l}_5=3$.
We notice that the neutral pion mass is $M_{\pi^0}^2$
very large already at $T=0$ due the strong magnetic field
$|qB|=5(140{\rm MeV})^2$.

\begin{figure}[htb]
\begin{center}
\setlength{\unitlength}{1mm}
{\includegraphics[width=8cm]{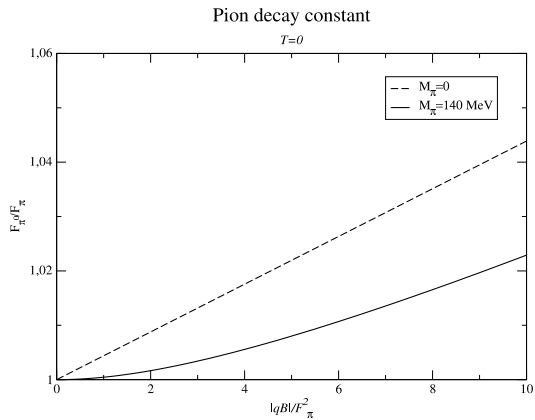}}
\caption{Pion decay constant  ${F}_{\pi^0}$ at $T=0$ scaled by 
its vacuum value $F_{\pi}$ as a function of $|qB|/F_{\pi}^2$.}
\label{fpivac}
\end{center}
\end{figure}

In Fig.~\ref{fpivac}, we show the pion decay constant
$F_{\pi}^0$ given by Eq.~(\ref{fpi00})
at $T=0$ normalized by its vacuum value $F_{\pi}=93$ MeV
as a function of $|qB|/F_{\pi}^2$ in the chiral limit as well as at the 
physical point. At $T=0$, the dependence on the magnetic field
is governed by the function $I_B(M)$ defined in Eq.~(\ref{ibdef}).
In the chiral limit, $F_{\pi^0}$ 
is a linear function of $|qB|$
since $I_B(0)=|qB|\log2$. In both cases, we note that
pion decays constant is an increasing function of $|qB|$.
We do not show $F_{\pi}^{\pm}$ since the functional dependence at $T=0$ is 
the same, cf. Eq.~(\ref{fpipm0}).

\begin{figure}[htb]
\begin{center}
\setlength{\unitlength}{1mm}
{\includegraphics[width=8cm]{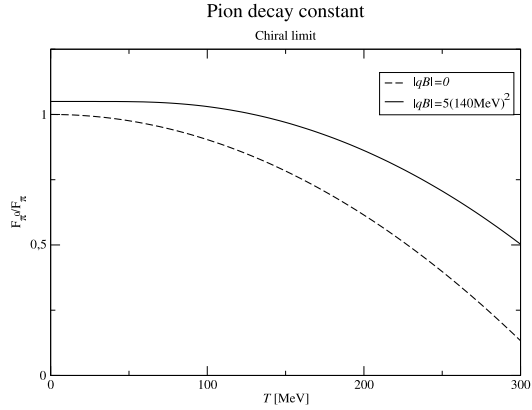}}
\caption{Normalized pion decay constant ${F}_{\pi^0}$ for 
and $|qB|=0$ and $|qB|=5(140 {\rm MeV})^2$ 
as a function of $T$.}
\label{fpib0}
\end{center}
\end{figure}

In Fig.~\ref{fpib0}, we show the normalized 
pion decay constant ${F}_{\pi^0}$ 
in the chiral limit
as a function of $T$ 
for $|qB|=0$ (dashed line)
and $|qB|=5(140 {\rm MeV})^2$ (solid line). 
In vanishing magnetic field, the temperature dependence is
given by 
\bqa
F_{\pi^0}=F_{\pi}\left(1-{T^2\over12F_{\pi}^2}\right)\;.
\eqa
The pion decay constant is larger in a magnetic background.
This is partly due to $I_B(M)$ which is an increasing function of $B$ and
partly due to the fact that $J_1^B|qB|\leq J_1T^2$ for all temperatures.

\begin{figure}[htb]
\begin{center}
\setlength{\unitlength}{1mm}
{\includegraphics[width=8cm]{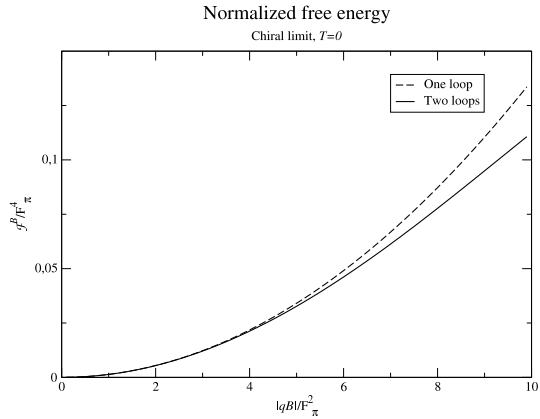}}
\caption{The contribution ${\cal F}^B$ 
to the one and two-loop free energy 
scaled by $F_{\pi}^4$
in the chiral limit.}
\label{free0}
\end{center}
\end{figure}

In Fig.~\ref{free0}, we show the contribution ${\cal F}^B$ to free 
energy 
in the chiral limit
at one and two loops normalized
to $F_{\pi}^4$ as a function of  $|qB|/F_{\pi}^2$. 
We have omitted the term ${(qB)^2\over6(4\pi)^2}\log{\Lambda^2\over2|qB|}$
as is customary in the literature. The uncertainty in the coupling
$\bar{d}=$ is large and even the sign is undetermined~\cite{belluci}.
Varying $\bar{d}$ would give rise to large bands for the
free energy as well as the quark condensate~\cite{werbos}.
Since the experimental value is consistent with zero, we will simply use
$\bar{d}=0$ in the remainder of the paper. 
The two-loop result~(\ref{f2bb}) for the free energy
then reduces to the one-loop result~(\ref{f1bb}) with an 
effective mass for the charged pion given by
$M_{\pi^{\pm}}^2={(qB)^2\over3(4\pi)^2F_{\pi}^2}(\bar{l}_6-\bar{l}_5)$.

\begin{figure}[htb]
\begin{center}
\setlength{\unitlength}{1mm}
{\includegraphics[width=8cm]{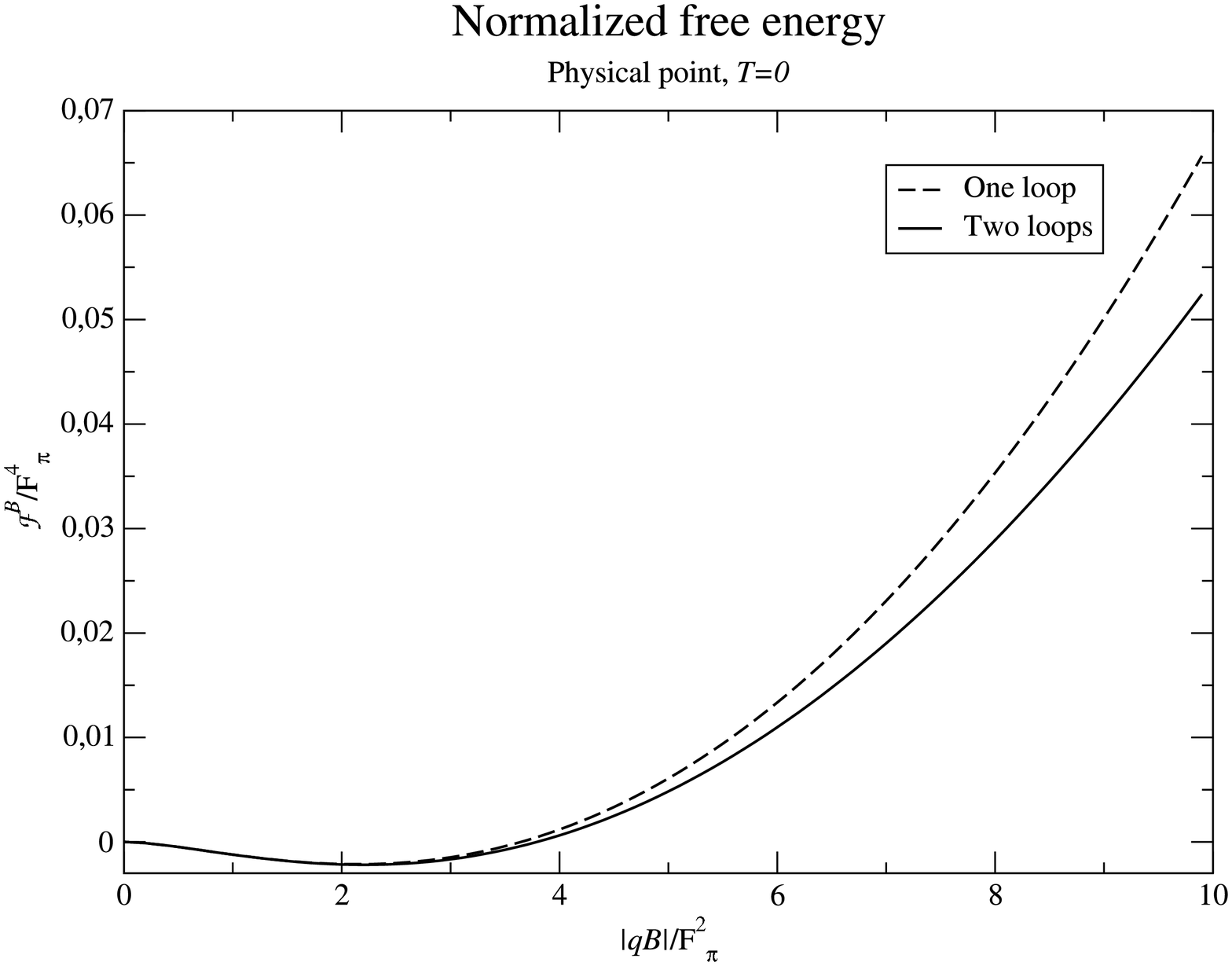}}
\caption{The contribution ${\cal F}^B$ to the one and two-loop free energy 
scaled by $F_{\pi}^4$
at the physical point.}
\label{freephys}
\end{center}
\end{figure}

In Fig.~\ref{freephys}, we show the contribution ${\cal F}^B$ to the
free energy 
at the physical point
at one - and two loops normalized
to $F_{\pi}^4$ as a function of $|qB|/F_{\pi}^2$. The correction is tiny and
the two-loop result starts to deviate from the one-loop result first
at very large values of the magnetic field.
Again the two-loop result is obtained from the one-loop formula~(\ref{f1bb})
using the
effective mass 
$M_{\pi^{\pm}}^2=M_{\pi}^2+{(qB)^2\over3(4\pi)^2F_{\pi}^2}(\bar{l}_6-\bar{l}_5)$
instead of $M_{\pi}^2$.

\begin{figure}[htb]
\begin{center}
\setlength{\unitlength}{1mm}
{\includegraphics[width=8cm]{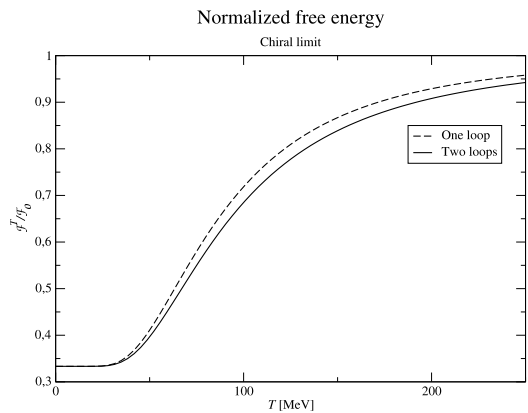}}
\caption{One and two-loop thermal part ${\cal F}^T$ of
the free energy normalized
to the free energy ${\cal F}_0=-\pi^2T^4/30$ of an ideal massless pion gas.}
\label{freech}
\end{center}
\end{figure}

In Fig.~\ref{freech}, we show the thermal part 
${\cal F}^T$ of the
free energy as a function of $T$ for $qB=5(140{\rm MeV})^2$
in the chiral limit normalized
to the free energy ${\cal F}_0=-\pi^2T^4/30$ of an ideal massless pion gas.
In the chiral limit, the two-loop result~(\ref{f222})
reduces to the one-loop~(\ref{1lv}) evaluated with the
$M^2_{\pi^{\pm}}=(qB)^2(\bar{l}_6-\bar{l}_5)/3(4\pi)^2F_{\pi}^2$
instead of $M_{\pi}=0$.
For low temperatures, only the neutral pion contributes
to the free energy since the contributions from the 
charged pions are Boltzmann suppressed. From $T$ around 50 MeV
onwards, the
normalized free energy increases sharply and 
approaches the normalized Stefan-Boltzmann result 1 as $T\rightarrow\infty$.
It has the value of approximately
0.97 for $T=313$ MeV, i. e. for $T^2/|qB|=1$.
The two-loop result lies below the one-loop result since 
in this approximation the charged pions are not massless but
has an effective mass due to the magnetic field,
$M^2_{\pi^{\pm}}=(qB)^2(\bar{l}_6-\bar{l}_5)/3(4\pi)^2F_{\pi}^2$.

\begin{figure}[htb]
\begin{center}
\includegraphics[width=8.cm]{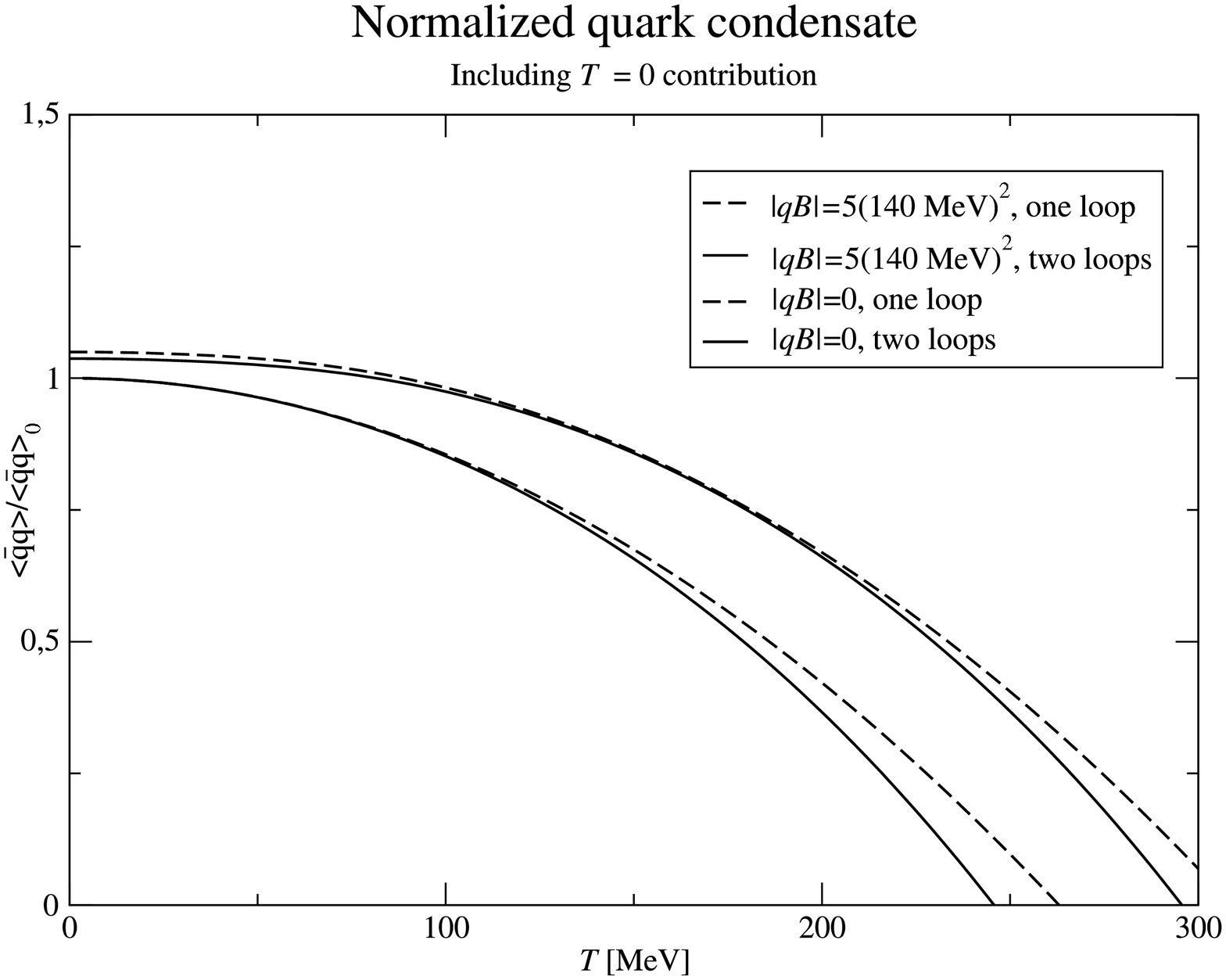}
\caption{
Temperature dependence of the quark condensate 
including the $T=0$ contribution
normalized to its vacuum value 
$|qB|=5\,\,(140\,\,{\rm MeV})^2$ at LO and NLO in chiral perturbation
theory.
For comparison, we show the LO and NLO results for $|qB|=0$ as well.}
\label{condensate}
\end{center}
\end{figure}

\begin{figure}[htb]
\begin{center}
\includegraphics[width=8.cm]{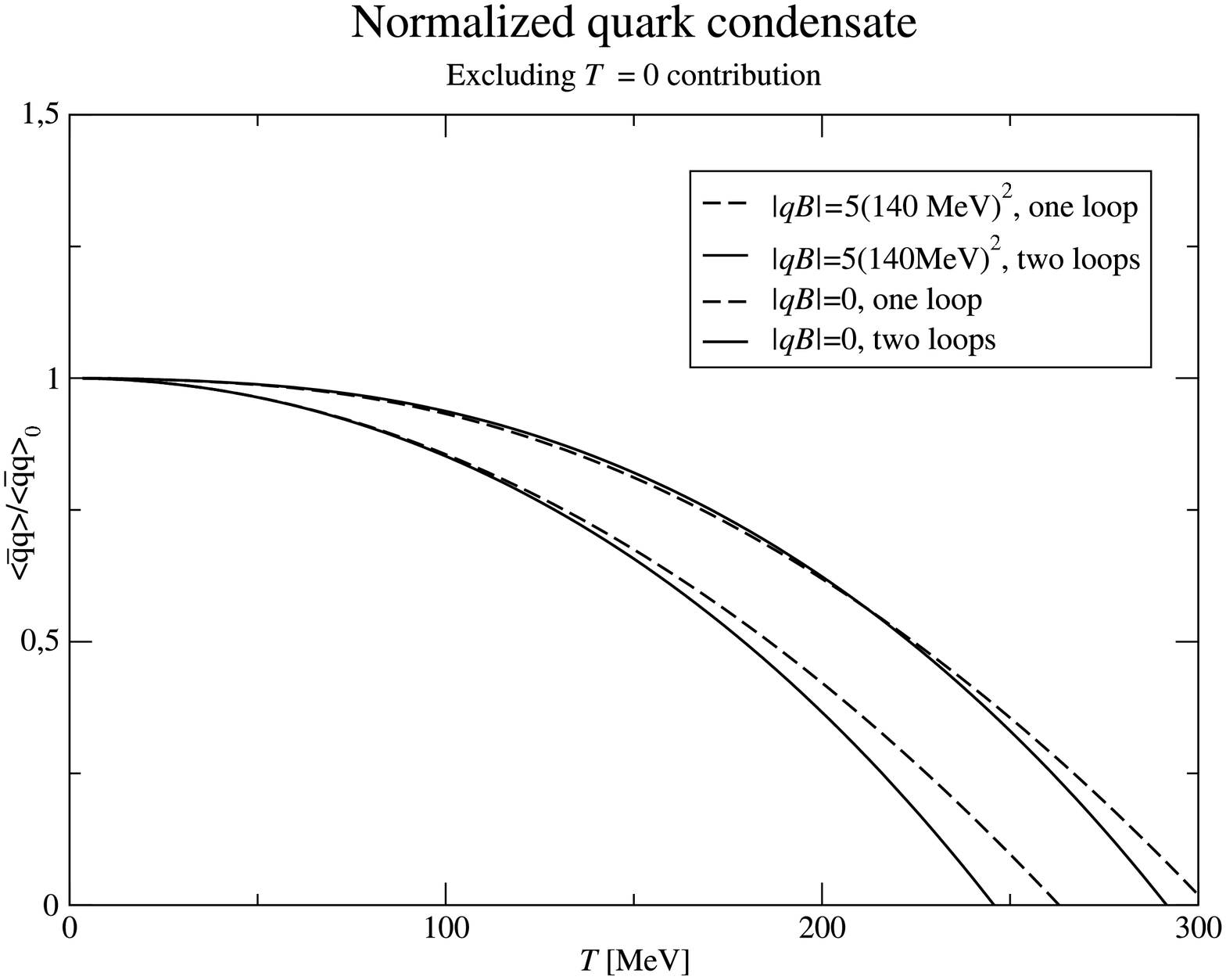}
\caption{
Temperature dependence of the quark condensate 
excluding the $T=0$ contribution
normalized to its vacuum value 
$|qB|=5\,\,(140\,\,{\rm MeV})^2$ at LO and NLO in chiral perturbation
theory.
For comparison, we show the LO and NLO results for $|qB|=0$ as well.}
\label{condensate2}
\end{center}
\end{figure}

In Fig.~\ref{condensate}, we show the LO and NLO
results for the 
normalized quark condensate
in the chiral limit including the $T=0$ contribution.
For comparison, we also show the LO and NLO results for the
quark condensate for $B=0$ given by Eq.~(\ref{condb0}).
In Fig.~\ref{condensate2}, we show the LO and NLO
results for the normalized quark condensate
in the chiral limit excluding the zero-temperature contributions in order
to disentangle the effects of the magnetic field at $T=0$ and the
finite-temperature effects.
For comparison, we again show the LO and NLO results for the
quark condensate for $B=0$.
Figs.~\ref{condensate} and~\ref{condensate2} show that the 
one-loop and two-loop results for the condensate
are very close to each other in the entire temperature
range. The LO and NLO results for the condensate are closer
in a magnetic background than for $B=0$ and this suggests that
chiral perturbation theory converges at least as well for
finite $B$. 

Figs.~\ref{condensate} and~\ref{condensate2} show that the 
quark condensate 
for nonzero $B$ goes to zero slower than for vanishing magnetic field
and the reason 
is two-fold. Firstly, there is an enhanced quark condensate at $T=0$, which
to leading order in ChPT is given by the function $I_B(M)$.
This is the familiar enhancement of the chiral condensate in a magnetic
background. Secondly, there are finite-temperature effects.
The basic mechanism is that $J_1^B$ is a decreasing function of 
the magnetic field $B$ and so $J_1T^2\geq J_1^B|qB|$ for all nonzero $B$.

Comparing the results for the condensate at $B=0$ and 
$|qB|=5(140 {\rm MeV})^2$, we see that its effects are quantitavely
large. This is a due to a very strong magnetic field.
For weaker magnetic fields, the difference between the two sets of 
curves will be smaller too. The results suggest that the
critical temperature $T_c$ for the chiral transition is higher in a magnetic
background. However, since ChPT is known to break down for large temperatures
we do not make any quantitative statements about $T_c$ as a function
of the magnetic field $B$.
In Ref.~\cite{chiralB}, the authors are using chiral perturbation theory
to investigate the quark-hadron 
phase transition as a function of the magnetic field at the physical point.
They compare the pressure of a hot pion gas with that of an ideal 
quark-gluon
plasma with with a subtracted vacuum energy term 
due to a nonzero gluon condensate.
For weak magnetic fields, the transition is first order. The line
of first-order transitions ends at critical point
$(\sqrt{|qB|},T)=(600,104)$ MeV. The critical temperature determined
this way is a decreasing function of the magnetic field $B$.

D'Elia {\it et al} have recently carried out lattice simulations in a constant
magnetic background~\cite{sanf,negro}.
They explored various constituent quark masses corresponding to a pion mass
of $200-480$ MeV and different magnetic fields, up to $|qB|\sim20$ $M_{\pi}^2$
for the lightest quark masses.
For these values of the pion mass, 
they found that
there is a slight increase in the critical temperature $T_c$ for the
chiral transition. These results have been confirmed by
Bali {\it et al}~\cite{budaleik,gunnar}.
The same group has alo carried out 
lattice simulations for physical values of the pion mass, 
i.e. $M_{\pi}=140$ MeV. Their results which are extrapolated to the continuum
limit show that the 
critical temperature is a decreasing function of the magnetic field
\cite{budaleik,gunnar}. Hence the critical temperature 
for fixed $|qB|$ as a function
of the quark mass is nontrivial. 
This is in stark contrast to most model calculations
that imply an increasing critical temperature as a functions of $B$.
These include mean-field calculations as well as functional
renormalization group calculations that incorporate quantum as well
as thermal fluctuations~\cite{anders,skokov}.
Since this discrepancy is not understood, more work is needed to
resolve this problem.

\section{Acknowledgments}
The author would like to thank Gunnar Bali,
G. Endrodi, and F. Bruckmann
for useful discussions on their lattice simulations.
He thanks the Niels Bohr International Academy and the Discovery Center for 
kind hospitality during the course of this work.
\appendix
\section{Sum-integrals}
In the imaginary-time formalism for thermal field theory, 
a boson has 
Euclidean 4-momentum $P=(P_0,{\bf p})$ with $P^2=P_0^2+{\bf p}^2$. 
The Euclidean energy $P_0$ has discrete values:
$P_0=2n\pi T$,
where $n$ is an integer. 
Loop diagrams involve sums over $P_0$ and integrals over ${\bf p}$. 
With dimensional regularization, the integral is generalized
to $d = 3-2 \epsilon$ spatial dimensions.
We define the dimensionally regularized sum-integral by
\bqa
\sumint_P&=&
T\sum_{P_0=2\pi nT}\int_p
\;,
\eqa
where the the integral is defined 
\bqa
\int_p=
\left({e^{\gamma_E}\Lambda^2\over4\pi}\right)^{\epsilon}
\int{d^dp\over(2\pi)^d}\;,
\eqa
and $\Lambda$ 
is an arbitrary momentum scale. The factor
$(e^{\gamma_E} /4\pi)^{\epsilon}$ 
is introduced so that, after minimal subtraction of the 
poles in  $\epsilon$ due to ultraviolet
divergences, $\Lambda$ 
coincides with the renormalization scale of the $\overline{\rm MS}$ 
renormalization scheme.

In the case of particles with electric charge $q$ in a constant magnetic
field $B$,
the sum-integral is replaced by a sum over Matsubara 
frequencies $P_0=2\pi nT$,
a sum over Landau levels $m$, and an integral over momenta in $d-2=1-2\epsilon$
dimensions. We then define 
\bqa
\sumint_P^B&=&
{|qB|\over2\pi}\sum_{m=0}^{\infty}
\,T\sum_{P_0=2\pi n T}
\int_{p_z}\;,
\eqa
where the the integral is defined 
\bqa
\int_{p_z}=
\left({e^{\gamma_E}\Lambda^2\over4\pi}\right)^{\epsilon}
\int{d^{d-2}p\over(2\pi)^{d-2}}\;,
\eqa
and where the prefactor ${|qB|\over2\pi}$ takes into account the degeneracy
of the Landau levels.

The specific sum-integrals that we need are
\bqa
\sumint_P\log\left[{P^2+M^2}\right]
&=&-{1\over2(4\pi)^2}\left({\Lambda^2\over M^2}\right)^{\epsilon}
\left[
\left({1\over\epsilon}+{3\over2}+{21+\pi^2\over12}\epsilon\right)M^4
+ 2J_0T^4+{\cal O}(\epsilon^2)
\right]\;,
\label{sigdiv} \\
\nonumber
\sumint_P^B\log\left[P^2_0+p_z^2+M_B^2\right]
&=&
{1\over2(4\pi)^2}
\left({\Lambda^2\over2|qB|}\right)^{\epsilon}
\left[
\left({(qB)^2\over3}-M^4
\right)\left(
{1\over\epsilon}+1
\right)
+8\zeta^{(1,0)}(-1,\mbox{$1\over2$}+x)(qB)^2
\right.\\ &&\left.
-2J_0^B|qB|T^2
+{\cal O}(\epsilon)
\right]\;,
\label{div1}
\\ 
\sumint_P{1\over P^2+M^2}
&=&-{1\over(4\pi)^2}\left({\Lambda^2\over M^2}\right)^{\epsilon}
\bigg\{
\left[
{1\over\epsilon}+1
+\left({\pi^2\over12}+1\right)\epsilon\right]M^2
-J_1T^2+{\cal O}(\epsilon^2)
\bigg\}\;,
\label{div2}
\\ \nonumber
\sumint_P^B{1\over P_0^2+p_z^2+M_B^2}
&=&
-{1\over(4\pi)^2}\left({\Lambda^2\over2|qB|}\right)^{\epsilon}
\bigg\{
\left({1\over\epsilon}+{\pi^2\over12}\epsilon\right)M^2
-2\zeta^{(1,0)}(0,\mbox{$1\over2$}+x)|qB|
\\&&
-\zeta^{(2,0)}(0,\mbox{$1\over2$}+x)|qB|\,\epsilon
-J_1^B|qB|
+{\cal O}(\epsilon^2)
\bigg\}\;,
\label{div3}
\eqa
where $M_B^2=M^2+(2m+1)|qB|$ and where
we have defined the functions
\bqa
\label{jn}
J_n(\beta M)&=&
{4e^{\gamma_E\epsilon}\Gamma({1\over2})\over\Gamma({5\over2}-n-\epsilon)}
\beta^{4-2n}M^{2\epsilon}
\int_0^{\infty}
dp{p^{4-2n-2\epsilon}\over\sqrt{p^2+M^2}}{1\over e^{\beta\sqrt{p^2+M^2}}-1}\;,
\\ 
J_n^B(\beta M)&=&
{8e^{\gamma_E\epsilon}\Gamma({1\over2})\over\Gamma({3\over2}-n-\epsilon)}\beta^{2-2n}(2|qB|)^{2\epsilon}
\sum_{m=0}^{\infty}
\int_0^{\infty}
dp{p^{2-2n-2\epsilon}\over\sqrt{p^2+M_B^2}}{1\over e^{\beta\sqrt{p^2+M_B^2}}-1}
\;.
\label{jbn}
\eqa
The functions $J_n(x)$ and $J_n^B(x)$ satisfy the recursion relations
\bqa
\label{rec1}
xJ^{\prime}_n(x)&=&2\epsilon J_n(x)-2x^2J_{n+1}(x)\;,\\
xJ^{B\prime}_n(x)&=&-2x^2J^B_{n+1}(x)\;.
\label{rec2}
\eqa

\section{Explicit calculations}
In this appendix, we explitcitly calculate the sum-integrals that we need.

The first sum-integral we need is given by Eq.~(\ref{sigdiv}).
Summing over Matsubara frequencies we can write
\bqa
\sumint_P\log\left[{P^2+M^2}\right]
&=&\int_p\bigg\{\sqrt{p^2+M^2} 
+2T\log\left[1-e^{-\beta\sqrt{p^2+M^2}}
\right]\bigg\}
\;.
\label{divi}
\eqa
The first term in Eq.~(\ref{divi}) 
is ultraviolet divergent. Calculating it with dimensional 
regularization, we obtain
\bqa
\int_p
\sqrt{p^2+M^2}
&=&
-\left({e^{\gamma_E}\Lambda^2\over4\pi}\right)^{\epsilon}
{\Gamma\left(-{d+1\over2}\right)\over(4\pi)^{{d+1\over2}}}M^{d+1}\;.
\label{dimmie}
\eqa
The second term in Eq.~(\ref{divi})
can be expressed 
in terms of $J_0$ defined in Eq.~(\ref{jn})
using integration by parts.
Expanding Eq.~(\ref{dimmie}) in powers of $\epsilon$
through order $\epsilon$, we obtain Eq.~(\ref{sigdiv}).
We next consider the 
sum-integral~(\ref{div1}).
Summing over the Matsubara frequencies, we obtain
\bqa
\sumint_P^B\log\left[{P_0^2+p_z^2+M_B^2}\right]&=&
{|qB|\over2\pi}\sum_{m=0}^{\infty}
\int_{p_z}\left\{
\sqrt{p_z^2+M^2_B}
+2T\log\left[1-e^{-\beta\sqrt{p^2+M^2_B}}\right]\right\}
\;.
\label{summing}
\eqa
The first integral is ultraviolet divergent and we compute it
in dimensional
regularization using Eq.~(\ref{dimmie}) with $d=1-2\epsilon$. This yields
\bqa
\int_{p_z}\sqrt{p_z^2+M^2_B}
&=&
-{1\over4\pi}
\left({e^{\gamma_E}\Lambda^2\over M^2_B}\right)^{\epsilon}
\Gamma(-1+\epsilon)M_B^2\;.
\label{10e}
\eqa
Eq.~(\ref{10e}) shows that the sum over Landau levels $m$
involves the term $M_B^{2-2\epsilon}$. This sum is divergent for $\epsilon=0$
and we regulate it using zeta-function regularization.
After scaling out a factor
of $(2|qB|)^{1-\epsilon}$, this sum can be written as
\bqa\nonumber
\sum_{m=0}^{\infty}
M_B^{2-2\epsilon}&=&
(2|qB|)^{1-\epsilon}
\sum_{m=0}^{\infty}
\left[m+\mbox{$1\over2$}
+{M^2\over2|qB|}
\right]^{1-\epsilon}\\
&=&
(2|qB|)^{1-\epsilon}
\zeta(-1+\epsilon,\mbox{$1\over2$}+x)\;,
\eqa
where $x={M^2\over2|qB|}$ and $\zeta(q,s)=\sum_{m=0}^{\infty}(q+m)^{-s}$ 
is the Hurwitz zeta-function.
We then find
\bqa
{|qB|\over2\pi}\sum_{m=0}^{\infty}
\int_{p_z}\left\{
\sqrt{p_z^2+M^2_B}
\right\}
&=&
-{4\over(4\pi)^2}
\left({e^{\gamma_E}\Lambda^2\over2|qB|}\right)^{\epsilon}
\Gamma(-1+\epsilon)
\zeta(-1+\epsilon,\mbox{$1\over2$}+x)(qB)^2\;.
\label{vakleik}
\eqa
The second term in Eq.~(\ref{summing})
can be expressed 
in terms of $J_0^B$ defined in Eq.~(\ref{jbn})
using integration by parts.
Expanding Eq.~(\ref{vakleik}) in powers of $\epsilon$, we obtain
Eq.~(\ref{div1}).


Taking the derivative of Eq.~(\ref{sigdiv}) with respect to $M^2$
and using the recursion relation~(\ref{rec1}) for $J_n$, one obtains
Eq.~(\ref{div2}).

We next consider Eq.~(\ref{div3}). It is given by
the derivative of Eq.~(\ref{div1}) with respect to $M_B^2$.
The temperature-independent part
of the sum-integral is given by the derivative of Eq.~(\ref{vakleik})
with respect to $M^2$.
This yields
\bqa
{|qB|\over2\pi}\sum_{m=0}^{\infty}
\int_{p_z}
{1\over2\sqrt{p_z^2+M^2_B}}
&=&
{2|qB|\over(4\pi)^2}
\left({e^{\gamma_E}\Lambda^2\over2|qB|}\right)^{\epsilon}
\Gamma(\epsilon)
\zeta(\epsilon,\mbox{$1\over2$}+x)\;.
\label{d1}
\eqa
Using the recursion relation~(\ref{rec2}), the temperature-dependent
term of Eq.~(\ref{div3}) is expressed in terms of $J_1^B$.

\end{document}